\documentclass[11pt,onecolumn]{article}
\usepackage[margin=1in]{geometry}
\usepackage{amsmath,amssymb}
\usepackage{graphicx}
\usepackage{booktabs}
\usepackage{array}
\usepackage[numbers,sort&compress]{natbib}
\usepackage{url}
\usepackage[colorlinks=true,allcolors=blue]{hyperref}
\hypersetup{
  pdfauthor={Akın Oktav},
  pdftitle={What does the model actually see? Evaluation protocols and input availability in data-driven prediction of room acoustic parameters},
  pdfsubject={Accepted manuscript. J. Acoust. Soc. Am. 160(5), 4251-4262 (2026), https://doi.org/10.1121/10.0047539},
  pdfkeywords={room acoustics; machine learning; evaluation protocols; spatial cross-validation; ISO 3382-1}
}
\usepackage{authblk}
\DeclareGraphicsExtensions{.pdf,.png,.jpg,.jpeg}

\title{What does the model actually see? Evaluation protocols and input availability in data-driven prediction of room acoustic parameters}
\author[ ]{Ak\i n Oktav}
\affil[ ]{Vibration and Acoustics Laboratory (VAL) and Department of Mechanical Engineering, Alanya Alaaddin Keykubat University, 07425, Antalya, T\"urkiye}
\date{\vspace{-1.2em}\normalsize\texttt{akin.oktav@alanya.edu.tr}}

\begin{document}
\maketitle

\begin{abstract}
Machine-learnt models are increasingly used to predict ISO 3382-1 room
acoustic parameters at unmeasured seats from
sparse measurements, with reported
coefficients of determination frequently above 0.85. This paper shows
that such figures are often determined by the evaluation protocol
rather than by the model. Using a multi-condition measurement campaign
in a 264-seat conference hall and a 180-seat concert hall, three model
families were evaluated under a factorial protocol ablation: validation
splits either row-based or grouped by receiver position, and inputs
either including measured-at-test quantities (the target position's
impulse response, co-located parameter measurements, position
identifiers) or restricted to source--receiver geometry and
environmental state. Row-based splits with measured-at-test inputs
reproduce the high reported accuracies (mean $R^2$ of 0.81 for the core
parameters); grouped splits with deployment-consistent inputs reduce
these to 0.09--0.60 and reorder the apparent difficulty of parameter
classes. Access to the target's own impulse response at test time reproduces the high accuracy of the row-based folds but provides no benefit under grouped folds. The row-based figure reflects condition interpolation at the measured positions rather than transferable acoustic information. Under the deployment-consistent
protocol the spread between Random Forest, the hybrid network, and
inverse-distance weighting is several times smaller than the
spread between protocols for a fixed model; the learnt models retain a
measurable advantage for sound strength and reverberation time, and the
high accuracy of the original pipelines re-emerges as condition
interpolation at measured positions, a distinct and operationally
useful task. A reporting checklist operationalises the findings.
\end{abstract}

\noindent\fbox{\parbox{\dimexpr\linewidth-2\fboxsep-2\fboxrule}{\small
This is the accepted manuscript of an article published in J.~Acoust.\ Soc.\ Am.\ \textbf{160}(5), 4251--4262 (2026), \url{https://doi.org/10.1121/10.0047539}. The published version differs in copy-editing and typesetting. Copyright 2026 Acoustical Society of America. This article may be downloaded for personal use only. Any other use requires prior permission of the author and the Acoustical Society of America.}}
\medskip

\section{Introduction}
\label{sec:intro}

The spatial characterisation of a performance venue under ISO 3382-1
requires impulse responses to be measured at many receiver positions
distributed over the audience area~\cite{iso2009}. Because the cost of a
measurement campaign scales with the number of positions, there is a
long-standing interest in predicting acoustic parameters at positions
that have not been measured, using a sparse set of measured positions
together with geometric and environmental information. Neural networks were applied to this family of problems as early as the 1990s~\cite{nannariello1999,nannariello2001,nannariello2001b}, geometry-based learning has since been revisited with modern models~\cite{falconperez2018}, and the recent growth of data-driven acoustics~\cite{bianco2019} has produced a broad range of related models, from blind estimation of parameters from reverberant
signals~\cite{eaton2016,deng2020,duangpummet2022,zhang2024} to deep
networks operating on measured or simulated impulse
responses~\cite{yu2021,yeh2021,bakos2025}, surveyed
in~\cite{vanwaterschoot2025}. Reported accuracies are often high; in
the author's work, coefficients of determination above 0.85 have been
reported for several ISO 3382-1 parameters in individual
venues~\cite{companion}.

This paper examines a prior question without which such figures cannot be interpreted: what did the model actually see, both
during validation and at inference? A reported coefficient of
determination is the answer to whatever question the evaluation
protocol poses, and two protocol choices, usually made silently,
determine that question. The first is the design of the validation
split. The second is the set of inputs the model receives at inference
time, and specifically whether each input would exist at a genuinely
unmeasured position.

The pattern documented here is not unique to acoustics. In the
machine-learning literature the general phenomenon is known as
leakage, the use during training or validation of information that
would not be available at deployment~\cite{kaufman2012}, and it has
been identified as a leading cause of irreproducible accuracy claims
across at least seventeen scientific fields~\cite{kapoor2023}. In
clinical applications, record-wise cross-validation of repeated
measurements from the same subjects inflates accuracy relative to
subject-wise validation~\cite{saeb2017}. In ecology and geostatistics,
random cross-validation of spatially autocorrelated observations
overstates the skill of mapping models, motivating spatially blocked
designs~\cite{roberts2017}, and spatial validation of large-scale
ecological maps has revealed sharp drops from their randomly validated
figures~\cite{ploton2020}. Room-acoustic measurement campaigns share
both structures in a statistical sense: multiple rows share a single nominal receiver position (measured from different sources or under different conditions), and adjacent positions are not independent of one another, at scales that vary with the quantity and the frequency band. These similarities form the basis of the validation design; they imply neither a common physical correlation model nor the repeated measurement of a fixed measurand. This paper transfers the
blocked-validation insight to room acoustics and quantifies its
consequences for the ISO 3382-1 parameter set.

Both choices are consequential in room acoustics for a structural
reason: measurement campaigns produce many observations per receiver
position. A single position is typically measured from several source
positions and under several environmental states, so a campaign with
tens of positions yields hundreds of rows. When validation folds are
drawn at random over rows, the same nominal receiver position appears in both
training and test sets measured from a different source or under a different environmental state, and the
protocol measures interpolation across repeated conditions at measured
positions rather than prediction at unmeasured ones. This paper refers to the
former as condition interpolation and to the latter as spatial
prediction; both are legitimate capabilities, but they are different
claims, and only grouped validation, in which all rows belonging to a
held-out position are excluded from training, tests the second. The
input set raises a parallel distinction. Features such as the target
position's own impulse response, measured values of related parameters
at the same position, or the position identifier itself travel with the
test rows and are unavailable at an unmeasured seat. The point is
sharpest for the parameters derived from the same omnidirectional impulse response (T30, EDT, Ts, C80, D50): each of them is
computed from that one impulse response, so supplying one of them as an input for another presupposes that the target position has
already been measured.

Data
quality is the complementary precondition: a separate audit study
quantified how field-campaign errors, from receiver-label swaps to
level-calibration interruptions, propagate into learnt spatial
models~\cite{nvw2026}; the present paper concerns the validity of the
evaluation protocol even when the data themselves are free of such
errors. The author's previous protocols are included in this observation: the analyses
that motivated this paper were carried out on the author's in-press and submitted results.

The contribution of this paper is a controlled, factorial dissection of
these protocol effects on a single multi-condition measurement
campaign, conducted in two halls of contrasting design, a 180-seat
shoebox concert hall and a 264-seat stepped conference hall, and
analysed with three model families: a Random Forest regressor over
engineered geometric and environmental features, a hybrid convolutional
network that fuses a geometric branch with a mel-spectrogram branch
through learnable weights, and simple spatial baselines
(nearest-neighbour and inverse-distance weighting). Specifically, the paper
(i) quantifies, by a two-way ablation over split design and feature
availability, how much of the reported accuracy of an existing pipeline
is attributable to each protocol choice; (ii) demonstrates that access to the target's own impulse response at test time, which yields the high row-based accuracy, does not generalise to previously unseen positions in a venue with few measured positions, so that the row-based figure reflects condition interpolation rather than spatial prediction; (iii) demonstrates that under
the deployment-consistent protocol, with grouped splits and geometry-plus-environment
inputs only, the three model families converge to comparable and modest
accuracy, which reorders the apparent difficulty of parameter classes;
and (iv) re-derives spatial-sampling (measurement-density) guidance from a held-out-position
design, in which models are trained on a subset of positions and
evaluated on the excluded ones, leading to a short reporting checklist
for data-driven room acoustics.

The remainder of the paper is organised as follows.
Section~\ref{sec:campaign} summarises the measurement campaign and the
parameter set, with full details in~\cite{companion}.
Section~\ref{sec:taxonomy} develops the protocol taxonomy.
Section~\ref{sec:experiments} reports the factorial ablation, the
test-time signal-access analysis, the cross-model comparison, and the
held-out-position density curves. Section~\ref{sec:discussion}
discusses the physical reading of the reordered parameter difficulty
and the legitimate uses of condition interpolation, and
Section~\ref{sec:checklist} condenses the findings into a reporting
checklist.

\section{Measurement campaign and parameter set}
\label{sec:campaign}

\begin{figure}[t]
\centering
\includegraphics[width=\linewidth]{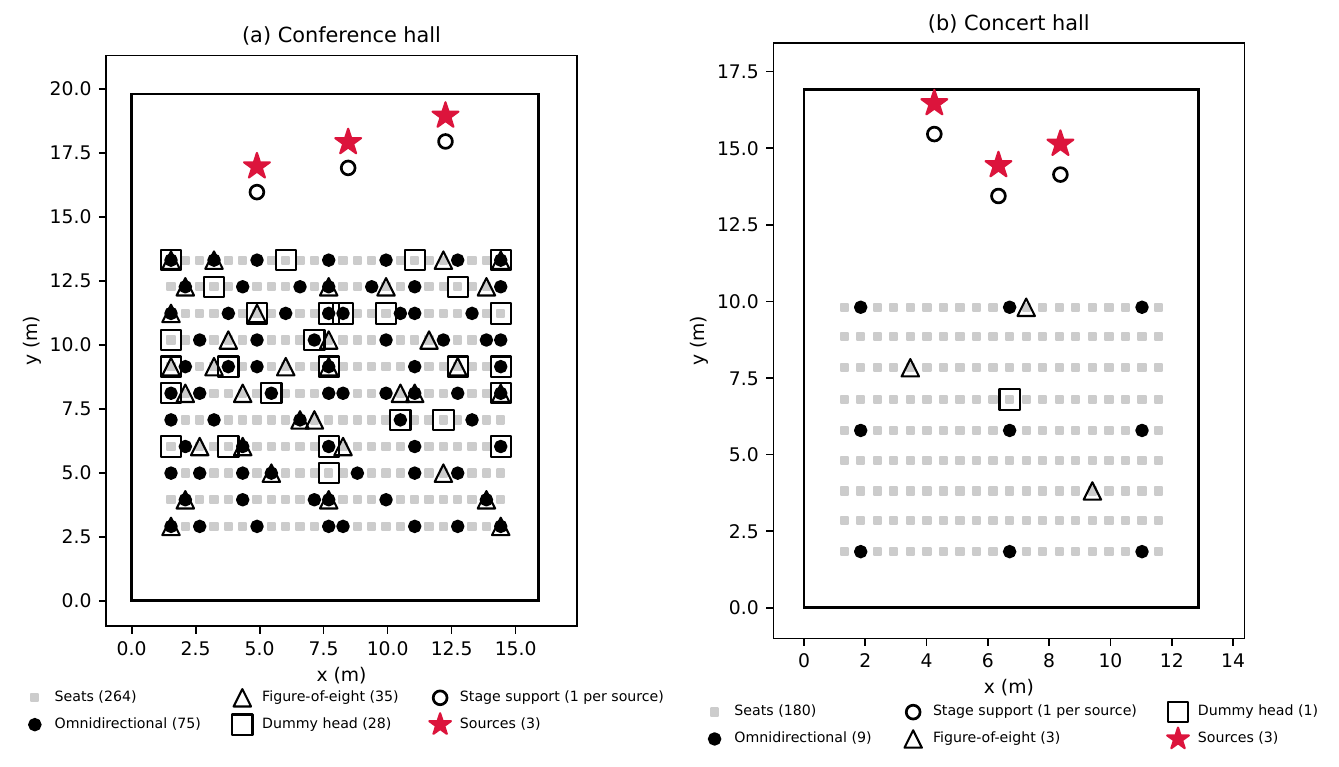}
\caption{Source and receiver positions. (a)~Conference hall: 75 omnidirectional receivers over the 264-seat audience area, the 35 figure-of-eight and 28 dummy-head positions, and the three stage sources. (b)~Concert hall: nine omnidirectional audience receivers on a $3\times3$ grid (rows 1, 5 and 9; seats 2, 11 and 19), and the figure-of-eight, dummy-head and source positions. Stage-support microphones stood 1\,m in front of each source in both halls; only the concert-hall stage-support rows are part of the analysed dataset. Dimensions in metres, from the surveyed seat coordinates.}
\label{fig:positions}
\end{figure}

All experiments in this paper re-analyse a single multi-condition
measurement campaign, summarised here in the detail needed to make the
paper self-contained; the companion study~\cite{companion} reports the
spatial-density analyses of the same campaign. Two halls of contrasting design were
measured under ISO 3382-1~\cite{iso2009}: the Koral \c{C}algan Concert
Hall (Anadolu University; shoebox geometry, 1538~m\textsuperscript{3},
180 seats) and the Alev Alatl{\i} Conference Hall (Alanya Alaaddin
Keykubat University; stepped multi-purpose geometry,
1077~m\textsuperscript{3}, 264 seats). In each hall, impulse responses
were recorded from three stage source positions at receiver positions
distributed over the audience area, in the empty hall at 24~$^{\circ}$C. Environmental variation was added at a subset of positions: in the conference hall, four positions were re-measured in ten empty-hall sessions spanning 21.0--25.5~$^{\circ}$C and three of them at 25\% to 100\% occupancy with seated listeners in regular and random arrangements, at the nominal 24~$^{\circ}$C set before the audience was admitted. In the concert hall, all ten positions were re-measured at 25\% and 50\% occupancy and four positions in ten sessions spanning 22.5--27.0~$^{\circ}$C. The impulse responses were recorded using a sine-sweep stimulus~\cite{farina2000}; T30 and EDT were determined from the energy decay curve, obtained by integrating the squared response backwards~\cite{schroeder1965}, whilst the remaining parameters were determined using their respective analysis methods based on the measured responses, as described below. The omnidirectional dataset used here
comprises 189 rows at ten omnidirectional receiver labels in the
concert hall, nine distributed over the audience area and one stage-support label whose microphone stood 1\,m in front of whichever source was active, so that its rows, which also carry strength and clarity values, form one group in the grouped folds although they occupy three physical points (the campaign additionally included three
figure-of-eight positions and one dummy-head position) and
285 rows at 75 receiver positions in the conference hall, so that
the conference hall carries the spatial analyses while the concert hall
provides a small-position contrast case. Ten room-acoustic quantities are considered (T30, EDT, ITDG, Ts, C80, D50, BR, LF, G, and a
binaural index), eight of them defined in ISO 3382-1, each at up to six octave bands (125~Hz to 4~kHz). These quantities do not share one measurement arrangement. T30, EDT, Ts, C80 and D50 are derived from the same omnidirectional impulse response, and its energy decay curve, at a given position and condition, so they inherently share information; G requires an absolute acoustic reference, LF a figure-of-eight receiver, and the binaural quantity a dummy-head measurement. Two descriptors are not defined in ISO 3382-1: the bass ratio BR is the ratio of the reverberation times at 125 and 250~Hz to those at 500~Hz and 1~kHz, and therefore shares the T30 information, and the initial time-delay gap ITDG is the interval between the direct sound and the first reflection at least 5~dB above the local noise floor, evaluated after the floor reflection, whose delay is set by the local source and receiver heights rather than by the room. The binaural index is the interaural cross-correlation coefficient IACC over 0--80~ms. Impulse responses were captured with a dodecahedral loudspeaker and
an omnidirectional condenser microphone at 1.2\,m above the local floor, that is, at absolute heights of 1.2--2.3\,m in the stepped conference hall and 1.3--2.4\,m in the raked concert hall, with
figure-of-eight and dummy-head microphones at the subset of positions
noted above (Figure~\ref{fig:positions}). The level calibration of the measurement chain was checked before each session. The reference level for sound strength was established once per hall, at the start of the block of consecutive empty-hall sessions, using the source calibration procedure provided by the measurement software (Dirac); no free-field measurement of the source at a distance of 10\,m was carried out, meaning that the G values for both halls refer to a fixed in-house reference rather than to the free-field reference specified in ISO 3382-1. A common offset leaves $R^2$ and all within-hall comparisons unchanged. The sessions with an audience in the conference hall show reference levels that deviate by 3 to 13\,dB from the empty-hall sessions at the same positions, and are excluded from the G analysis (242 out of 285 rows retained); in the concert hall, the session-mean G differences from the empty-hall survey at the same receiver and source are within 0.4\,dB, with one band-dependent deviation at a single position in one temperature session, which is a local anomaly rather than a reference shift.
In regular arrangements the seats were filled contiguously from the front rows; in random arrangements they were distributed uniformly over the seating area. Excitation
provided at least 35\,dB of broadband headroom above the background
level in both halls; the impulse-to-noise ratios of the concert-hall
measurements have a 125-Hz-band median of 55\,dB, and the small number
of rows failing the decay-range requirements of the low bands were
screened out prior to analysis. The closest audience source--receiver distances are 3.7\,m (conference hall) and 4.7\,m (concert hall). Across the ten empty-hall sessions, with the microphone left in place, the band values for the same source--receiver pair agree to within 0.2\,dB (G), 0.02\,s (T30), 0.3\,dB (C80) and 0.9\,ms (Ts) in the median across all bands and positions, and to within 0.5\,dB, 0.14\,s, 0.7\,dB and 4\,ms in the worst band and at the worst position (one standard deviation). The analysed G set for the conference hall thus comprises the survey session (212 rows, 72 positions, three sources) and the ten temperature sessions (30 rows at three of the four positions, one source).

Table~\ref{tab:measured} tabulates the mean, standard deviation and range of the measured quantities per hall and band.

\begin{table*}[t]
\centering
\caption{Measured quantities: mean (standard deviation) over all receiver positions and states, per hall and octave band; range over the full dataset in the last column. T30 and EDT in s, Ts and ITDG in ms, C80 and G in dB, D50, LF, IACC and BR dimensionless; BR and ITDG single-valued. G in both halls is relative to a fixed in-house reference (Section~\ref{sec:campaign}); LF and IACC from the conference-hall figure-of-eight and dummy-head positions.}
\label{tab:measured}
\scriptsize
\setlength{\tabcolsep}{2pt}
\begin{tabular}{lllllllll}
\hline
Hall & Par. & 125 & 250 & 500 & 1k & 2k & 4k & Range \\
\hline
Conf. & T30 & 0.81 (0.19) & 0.81 (0.10) & 0.67 (0.07) & 0.65 (0.07) & 0.73 (0.12) & 0.70 (0.10) & 0.18--1.81 \\
      & EDT & 0.65 (0.19) & 0.60 (0.14) & 0.52 (0.09) & 0.59 (0.10) & 0.59 (0.11) & 0.59 (0.11) & 0.18--1.38 \\
      & Ts  & 64 (14) & 41 (8) & 35 (6) & 35 (7) & 33 (6) & 36 (8) & 16--121 \\
      & C80 & 6.2 (2.6) & 8.6 (1.8) & 9.6 (1.4) & 8.9 (1.6) & 9.1 (1.4) & 8.6 (1.7) & $-0.8$--14.3 \\
      & D50 & 0.60 (0.16) & 0.75 (0.08) & 0.78 (0.06) & 0.76 (0.07) & 0.78 (0.05) & 0.75 (0.08) & 0.17--0.91 \\
      & G   & 10.5 (7.6) & 12.9 (7.3) & 12.4 (6.5) & 13.3 (6.5) & 15.0 (5.9) & 13.8 (6.1) & $-7.4$--26.6 \\
      & LF  & 0.09 (0.07) & 0.12 (0.08) & 0.16 (0.08) & 0.19 (0.09) & 0.13 (0.07) & 0.23 (0.13) & -- \\
      & IACC & 0.95 (0.03) & 0.80 (0.11) & 0.56 (0.10) & 0.49 (0.16) & 0.46 (0.16) & 0.34 (0.11) & -- \\
      & BR  & \multicolumn{6}{l}{1.28 (0.24)} & 0.70--2.60 \\
      & ITDG & \multicolumn{6}{l}{10.0 (4.4)} & 5.0--51.9 \\
\hline
Conc. & T30 & 0.95 (0.15) & 0.79 (0.05) & 0.68 (0.04) & 0.56 (0.05) & 0.46 (0.02) & 0.46 (0.03) & 0.38--2.40 \\
      & EDT & 0.82 (0.23) & 0.79 (0.15) & 0.61 (0.17) & 0.52 (0.10) & 0.43 (0.08) & 0.44 (0.09) & 0.00--1.46 \\
      & Ts  & 67 (20) & 52 (15) & 41 (12) & 30 (9) & 25 (8) & 27 (9) & 2--104 \\
      & C80 & 5.4 (3.2) & 6.1 (2.8) & 8.1 (3.0) & 10.4 (3.1) & 12.0 (2.5) & 11.7 (2.7) & $-0.3$--24.1 \\
      & D50 & 0.60 (0.16) & 0.66 (0.12) & 0.72 (0.10) & 0.79 (0.08) & 0.84 (0.06) & 0.83 (0.06) & 0.25--0.99 \\
      & G & $-11.0$ (4.1) & $-10.3$ (4.0) & $-10.5$ (4.2) & $-11.8$ (4.4) & $-12.5$ (4.6) & $-13.6$ (5.2) & $-20.0$--2.3 \\
\hline
\end{tabular}
\end{table*}

Measurement data quality for the campaign infrastructure was
screened using the framework introduced in~\cite{nvw2026}; the present analyses
therefore concern protocol validity on data screened for the field-error classes that framework addresses, not for measurement uncertainty in general, which is treated in Section~\ref{sec:limitations}. 

\section{A taxonomy of evaluation protocols}
\label{sec:taxonomy}

Throughout this paper, one observation (a row of the dataset) is a
measurement of one parameter at one receiver position $p$, from one
source position $s$, under one environmental state $e$ (a combination
of temperature, occupancy level, and seating arrangement). In
machine-learning terms, a \emph{fold} is one partition of the data
into a training and a test set; \emph{cross-validation} averages a
score over several such partitions; and a \emph{feature} is any input
column the model may read, and an \emph{ablation} removes one input or protocol element at a time to measure its effect. The physical question behind the
taxonomy is which acoustic measurements would have had to be
carried out at the target configuration for a given input to exist. A campaign
with $P$ receiver positions therefore yields many rows per position,
and the meaning of any cross-validated score depends on how the folds
treat this structure and on which columns the model is allowed to read.
These are the two axes of the taxonomy.

\subsection{Axis 1: split design and the unit of generalisation}
\label{sec:taxonomy-split}

A validation split implicitly declares a unit of generalisation. When
folds are drawn at random over rows, the unit is the individual
measurement: every receiver position contributes rows to both the
training and the test sets, differing only in source or environmental
state, and the protocol therefore estimates how well the model
interpolates across repeated conditions at positions it has already
seen. This is termed \emph{condition interpolation}. When folds are drawn
over receiver positions, so that every row belonging to a held-out
position is excluded from training, the unit is the position, and the
protocol estimates \emph{spatial prediction} at unmeasured receiver positions while the source configurations of the campaign remain represented in training. Prediction for a new receiver with represented sources is physically distinct from prediction for a new source or a previously unseen source--receiver pair; the claims of this paper are limited to the former, which is the case the sparse-measurement literature addresses.
Both estimates are legitimate, and both are operationally useful, but
they answer different questions and can differ by several tenths in the
coefficient of determination on the same data
(Section~\ref{sec:experiments}). Analogous groupings by source position
or by environmental state test complementary claims, such as robustness
to an unseen source location or an unvisited temperature, and can be
reported alongside; this paper focuses on the receiver-position grouping because the sparse-measurement literature phrases its claims in terms of unmeasured seats.

\subsection{Why the split design changes the measured accuracy}
\label{sec:theory}

The effect of the split design can be stated compactly. Let the
band-averaged value of a parameter at receiver position $p$ under
environmental condition $c$ be decomposed as
\begin{equation}
  y_{p,c} \;=\; \mu \;+\; s_p \;+\; e_c \;+\; \eta_{p,c},
  \label{eq:decomp}
\end{equation}
where $s_p$ is the time-invariant spatial component of the position,
$e_c$ the environmental component shared across positions,
$\eta_{p,c}$ the residual, and
$\sigma_s^2$, $\sigma_e^2$, $\sigma_\eta^2$ their variances. The source index is suppressed in Eq.~\eqref{eq:decomp} for illustration: $s_p$ denotes the receiver-associated component averaged over the represented sources, which is what receiver-grouped folds hold out. In a
multi-condition campaign every position is measured several times, from different sources or under different conditions, so any input channel that identifies the position, whether
an explicit identifier, coordinates acting as a lookup key, co-located
measurements of other parameters, or the position's own impulse
response, allows a flexible model to estimate $s_p$ directly from the
training rows of the same position. Under row-based validation such
rows are almost always present, and the apparent skill approaches
\begin{equation}
  R^2_{\mathrm{row}} \;\approx\;
  \frac{\sigma_s^{2} + \tilde{\sigma}_e^{2}}
       {\sigma_s^{2} + \sigma_e^{2} + \sigma_\eta^{2}},
  \label{eq:inflation}
\end{equation}
where $\tilde{\sigma}_e^{2}$ is the explainable part of the
environmental variance: the full spatial variance is credited to the
model even if none of it is predictable at a new position. Under
grouped validation the test positions contribute no training rows, the
memorisation channel is closed, and the credited spatial variance
reduces to the part of $s_p$ that is expressible in the available
inputs, for instance source--receiver geometry. The difference between
the two protocols is therefore governed by the fraction of
$\sigma_s^{2}$ that is memorisable but not transferable, and the
reordering of parameter classes observed in
Section~\ref{sec:ablation} follows: the clarity-class parameters, whose
spatial variance is dominated by seat-to-seat variation that the geometric descriptors do not resolve, show the largest inflation, while sound strength, whose
spatial component is largely captured by source--receiver distance, shows the
smallest. Equation~\eqref{eq:inflation} also explains why the
inflation persists for models with no explicit position identifiers:
any sufficiently rich position-correlated input reopens the channel.

\subsection{Axis 2: input availability at inference}
\label{sec:taxonomy-inputs}

The second axis classifies each model input by whether it would exist
at a genuinely unmeasured position. Four classes are distinguished.

\emph{(I1) The target position's own impulse response}, or a
representation of it such as a mel-spectrogram. By definition this
exists only where an impulse-response measurement has been made at the target seat. For the five parameters derived from the same omnidirectional response the point is structural rather than incidental: each of them is computed from the same impulse response, so a model
that receives I1 at inference is asked to reproduce quantities that are
already deterministic functions of its input.

\emph{(I2) Co-located measured quantities}: measured values of related
parameters at the same position (cross-parameter features) and
measurement-side descriptors such as the impulse-to-noise ratio or the
equivalent level. These likewise presuppose a completed parameter measurement at
the target position, and for the five omnidirectional-response targets they derive from
the same impulse response as the target itself; for LF, IACC and G the shared information is weaker, since these require a different receiver or an absolute reference.

\emph{(I3) Position identifiers}: the receiver or source index supplied
as a numeric or categorical feature. At a measured position an
identifier is an invitation to memorise; at an unmeasured position it
is an unseen symbol on which the model can only extrapolate.

\emph{(I4) Geometry and environmental state}: source and receiver
coordinates and their derived quantities, room dimensions, temperature,
occupancy, and seating arrangement. This is the only class that exists at an unmeasured position. Throughout, this paper calls the protocol that combines position-grouped folds with class-I4 inputs \emph{deployment-consistent} in the specific sense that every input consumed at inference would exist at the deployment target, an unmeasured seat.

Two further modes complete the picture. In the \emph{privileged} mode,
signal information (I1, and possibly I2) is used during training only,
while inference receives I4 alone; this is the learning-using-privileged-information setting of Vapnik and
Vashist~\cite{vapnik2009}, and it is the mode described by the author's
patent applications. It makes a deployment-consistent unmeasured-position claim,
because nothing unavailable is consumed at inference, and whether the
training-time signal actually improves the geometry-only predictor is
an empirical question that Section~\ref{sec:experiments} answers for
the present data. Finally, \emph{blind estimation} recovers parameters from a
signal recorded in situ at the position of
interest~\cite{eaton2016,zhang2024}; it legitimately consumes I1
because its claim concerns measured (or at least instrumented)
positions, and it is outside the scope of the unmeasured-position
question studied here.

\subsection{The protocol grid}
\label{sec:taxonomy-grid}

Crossing the two axes yields the grid of Table~\ref{tab:taxonomy}. Each
cell answers a different question, and one cell, grouped splits
with I4 inputs, corresponds to the question that sparse-measurement claims
pose. Grouped splits with I1 at test time form the diagnostic cell examined in Section~\ref{sec:cnn}.
The experiments that follow populate every cell of the grid on a single
campaign, so that the contribution of each protocol choice can be read
off directly.

\begin{table}[t]
\centering
\caption{Evaluation-protocol taxonomy: input class at inference (rows) against split design (columns). Each cell states the question the protocol answers; bold marks the cell that sparse-measurement (unmeasured-position) claims require.}
\label{tab:taxonomy}
\begin{tabular}{lll}
\hline
\parbox[t]{0.30\linewidth}{\raggedright Inference inputs\vspace{2pt}} & \parbox[t]{0.30\linewidth}{\raggedright Row-based folds\vspace{2pt}} & \parbox[t]{0.30\linewidth}{\raggedright Folds grouped by receiver position\vspace{2pt}} \\
\hline
\parbox[t]{0.30\linewidth}{\raggedright I1: target's own IR\vspace{2pt}} & \parbox[t]{0.30\linewidth}{\raggedright Parameter estimation from the position's own signal, evaluated across repeated conditions (near-tautological for the omnidirectional set)\vspace{2pt}} & \parbox[t]{0.30\linewidth}{\raggedright Diagnostic: does signal access transfer to unseen positions, or only identify training positions?\vspace{2pt}} \\
\parbox[t]{0.30\linewidth}{\raggedright I2: co-located measured parameters, measurement descriptors\vspace{2pt}} & \parbox[t]{0.30\linewidth}{\raggedright Condition interpolation with measured side-information\vspace{2pt}} & \parbox[t]{0.30\linewidth}{\raggedright Parameter completion at measured positions; presupposes measurement of the target position\vspace{2pt}} \\
\parbox[t]{0.30\linewidth}{\raggedright I3: position identifiers\vspace{2pt}} & \parbox[t]{0.30\linewidth}{\raggedright Memorisation-prone; inflates scores\vspace{2pt}} & \parbox[t]{0.30\linewidth}{\raggedright Extrapolation over unseen identifiers; typically degrades\vspace{2pt}} \\
\parbox[t]{0.30\linewidth}{\raggedright I4: geometry and environment\vspace{2pt}} & \parbox[t]{0.30\linewidth}{\raggedright Condition interpolation from geometry alone\vspace{2pt}} & \parbox[t]{0.30\linewidth}{\raggedright \textbf{Spatial prediction at unmeasured positions (the deployment-consistent cell)}\vspace{2pt}} \\
\parbox[t]{0.30\linewidth}{\raggedright Privileged: I1 in training only, I4 at inference\vspace{2pt}} & \parbox[t]{0.30\linewidth}{\raggedright (reduces to I4 row-based)\vspace{2pt}} & \parbox[t]{0.30\linewidth}{\raggedright Spatial prediction with training-time signal supervision~\cite{vapnik2009}\vspace{2pt}} \\
\hline

\end{tabular}
\end{table}

\section{Experiments}
\label{sec:experiments}

\subsection{Models, baselines, and common protocol}
\label{sec:models}

\begin{figure}[t]
\centering
\includegraphics[width=0.98\linewidth]{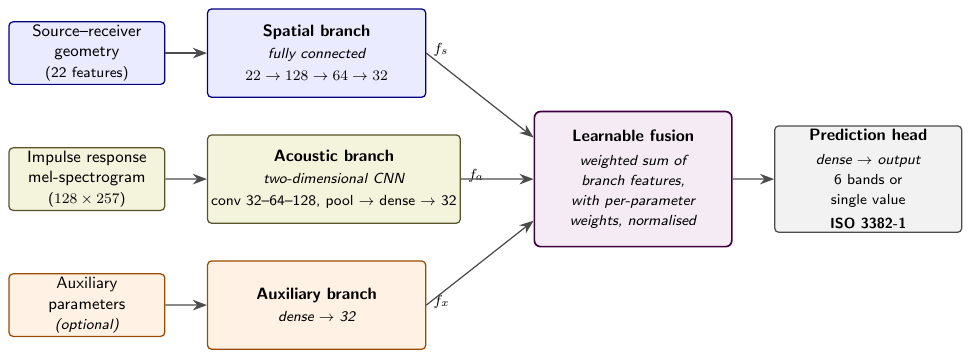}
\caption{Hybrid architecture. Solid paths are active in training and inference; dashed paths carry privileged inputs that exist only at measured positions and are used in training alone. At inference the acoustic and auxiliary branches are disabled and the prediction follows from the source--receiver geometry alone.}
\label{fig:arch}
\end{figure}

Three model families are evaluated. The first is the Random Forest~\cite{breiman2001} pipeline of the companion study~\cite{companion}: 150 trees, maximum
depth 12, an engineered feature set comprising seventeen spatial
features, three environmental features, and, in its reported configuration, receiver and source identifiers together with
cross-parameter and measurement-side auxiliaries. The second is a hybrid convolutional network (Figure~\ref{fig:arch}). A
fully connected spatial branch encodes twenty-two geometric and
environmental features (source and receiver coordinates, their relative
displacement, distance and angular terms, and the temperature,
occupancy, and seating state); a two-dimensional convolutional branch~\cite{lecun2015} encodes the mel-spectrogram of a measured impulse response (128 mel
bands, zero-padded to 257 frames), on the grounds that the features relevant to room
acoustic parameters, band-dependent decay slopes and the early-to-late
energy balance, are localised jointly in time and frequency. The two
feature vectors are combined by a weighted sum whose weights are learnt
jointly with the network and normalised to sum to one, so that each
parameter acquires its own spatial-acoustic balance; an optional
auxiliary branch admits co-located measured parameters where the
taxonomy permits them. The design element that matters for this paper
is the switchable inference mode: the acoustic and auxiliary branches
can be disabled at inference, with modality dropout during training
preparing the network for that mode, so that the input-availability
axis of Section~\ref{sec:taxonomy} is traversed without retraining.
This switchable, geometry-only inference mode is the element claimed in
the underlying patent application (TR 2026/001430). The third family comprises the simple spatial
baselines: nearest-neighbour assignment and inverse-distance weighting
(power 2) over the training positions, matched on source position.

Unless stated otherwise, grouped evaluations use five-fold
cross-validation grouped by receiver position, with internal validation
sets for early stopping likewise held out at the position level. Fold
compositions differ between the Random Forest and network harnesses,
but every grouped split excludes all rows of a held-out position from
training. The folds are not grouped by measurement session: rows from the same session appear on both sides of a split, and Section~\ref{sec:ablation} examines whether this is significant for G. Feature scaling is fitted inside each training fold. Two
aggregation conventions appear below and are not mixed within a table:
the Random Forest tables average per-band cross-validated $R^2$ over
octave bands, whereas the network tables pool the available bands into
a single $R^2$; cross-family comparisons are therefore read as
indicative rather than exact.

\subsection{Factorial protocol ablation (Random Forest, conference hall)}
\label{sec:ablation}

The companion pipeline was re-run under all six combinations of the two
protocol axes: split design (row-based KFold as reported, versus
GroupKFold by receiver position) and feature availability (FULL, the reported set including position identifiers, cross-parameter features,
and measurement-side auxiliaries; NO-ID, the same without position
identifiers; and GEO-ENV, geometry and environmental state only, the
class-I4 set of Section~\ref{sec:taxonomy}). Table~\ref{tab:ablation}
reports the six core parameters.

\begin{table}[t]
\centering
\caption{Factorial protocol ablation, conference hall (285 rows, 75 positions): cross-validated $R^2$ averaged over octave bands. ROW-FULL is the originally reported protocol; GROUP-GEO-ENV the deployment-consistent protocol of Table~\ref{tab:taxonomy}.}
\label{tab:ablation}
\footnotesize\setlength{\tabcolsep}{3pt}
\begin{tabular}{lcccccc}
\hline
 & \multicolumn{3}{c}{Row-based folds} & \multicolumn{3}{c}{Grouped by position} \\
Param. & FULL & NO-ID & GEO-ENV & FULL & NO-ID & GEO-ENV \\
\hline
G   & 0.856 & 0.842 & 0.778 & 0.639 & 0.660 & 0.604 \\
Ts  & 0.884 & 0.883 & 0.545 & 0.856 & 0.857 & 0.221 \\
C80 & 0.879 & 0.879 & 0.428 & 0.862 & 0.862 & 0.134 \\
D50 & 0.804 & 0.804 & 0.420 & 0.788 & 0.786 & 0.160 \\
EDT & 0.779 & 0.779 & 0.372 & 0.734 & 0.734 & 0.091 \\
T30 & 0.666 & 0.657 & 0.607 & 0.483 & 0.479 & 0.259 \\
\hline
\end{tabular}
\end{table}

Three attributions follow directly. First, the ROW-FULL column implements the companion study's protocol and feature set with the companion pipeline itself, which validates the ablation harness: all subsequent differences are attributable to the protocol, not to reimplementation. Second, for the clarity and temporal
parameters (Ts, C80, D50, EDT) the dominant term is feature
availability, not split design: grouping the folds barely moves the
FULL scores (C80: 0.879 to 0.862), because the cross-parameter inputs
travel with the test rows and carry the target information regardless
of how the folds are drawn, whereas removing the measured-at-test
inputs collapses the grouped scores to between 0.09 and 0.22. The
mechanism is the structural one of Section~\ref{sec:taxonomy}: the
cross-parameter inputs derive from the same impulse response as the
target. Third, position identifiers contribute essentially nothing under either design (G: 0.856 with and 0.842 without identifiers under row-based folds, 0.639 and 0.660 under grouped folds): the information they could carry is already present in the coordinate and distance features, so their presence in a feature set is a matter of reporting transparency rather than a source of accuracy. The deployment-consistent cell,
GROUP-GEO-ENV, leaves sound strength at 0.60, reverberation time at 0.26, and the clarity and temporal parameters between 0.09 and 0.22,
which reorders the apparent difficulty of parameter classes relative to the earlier reading. The two cells are juxtaposed per parameter in the first and last columns of Table~\ref{tab:ablation}; the differences, up to 0.75 for clarity, are the
paper's central quantity. Since the G rows for the conference hall were derived from eleven measurement sessions (Section~\ref{sec:campaign}), the ablation was repeated with G limited to the survey session alone (212 rows, 72 positions): the GROUP-GEO-ENV value increases by 0.17, and its fold-to-fold standard deviation decreases from 0.27 to 0.09 (both runs in the same computational environment), so that the session structure does not inflate the reported figure; the two positions that were measured only in the temperature block are the most difficult held-out cases.

\subsection{The hybrid network under grouped splits: privileged
information and test-time signal access}
\label{sec:cnn}

The network was evaluated on both halls for sound strength, clarity, reverberation time, and early decay time
under four conditions sharing the same grouped splits, alongside the
two baselines: K1, the spatial branch alone (class I4); K2, the full
hybrid trained with modality dropout (the acoustic branch disabled with
probability 0.5 during training) and evaluated with the acoustic branch
off, which realises the privileged mode of
Section~\ref{sec:taxonomy}~\cite{vapnik2009}; K2n, the hybrid trained
normally and ablated only at test time, a diagnostic for distribution
shift; and K3, the hybrid with the target position's own
mel-spectrogram supplied at test time, which is the originally reported protocol evaluated under grouped splits. Table~\ref{tab:cnn} reports pooled-band
$R^2$ for the four parameters.

\begin{table*}[t]
\centering
\caption{Hybrid-network conditions under grouped five-fold cross-validation (pooled bands, mean over folds). Row-based legacy: the originally reported row-based figure for the same model class (G). Fold-to-fold standard deviations are 0.1--0.4 in the conference hall and 0.7--1.2 for G in the concert hall (ten receiver groups; larger for K2n); per-fold values are released with the code. Pooled-band $R^2$ aggregates bands and samples jointly and, notably for T30, is buoyed by the band profile; it is not comparable to the band means of Table~\ref{tab:ablation}.}
\label{tab:cnn}
\small
\begin{tabular}{lcccccccc}
\hline
 & \multicolumn{4}{c}{Concert hall} & \multicolumn{4}{c}{Conference hall} \\
Condition & G & C80 & T30 & EDT & G & C80 & T30 & EDT \\
\hline
K1: geometry only (I4)  & $+0.19$ & $-0.12$ & $+0.69$ & $+0.45$ & $+0.60$ & $+0.25$ & $+0.46$ & $+0.20$ \\
K2: privileged (train-only signal)  & $-0.12$ & $+0.03$ & $+0.65$ & $+0.23$ & $+0.54$ & $+0.21$ & $+0.46$ & $+0.20$ \\
K2n: naive test-time ablation  & $-1.94$ & $-0.01$ & $-0.24$ & $-0.61$ & $-0.04$ & $-0.52$ & $-0.79$ & $-0.76$ \\
K3: target RIR at test (I1)  & $-0.09$ & $+0.23$ & $+0.77$ & $+0.39$ & $+0.68$ & $+0.33$ & $+0.58$ & $+0.17$ \\
Nearest neighbour  & $-1.10$ & $+0.08$ & $+0.83$ & $+0.11$ & $+0.24$ & $+0.08$ & $+0.04$ & $-0.24$ \\
Inverse-distance weighting  & $-0.21$ & $+0.40$ & $+0.86$ & $+0.42$ & $+0.53$ & $+0.43$ & $+0.34$ & $+0.23$ \\
\hline
Row-based legacy (I1, row folds)   & $0.93$  & --      & --      & --      & $0.85$  & --      & --      & --      \\
\hline
\end{tabular}
\end{table*}

\begin{figure}[t]
\centering
\includegraphics[width=0.92\linewidth]{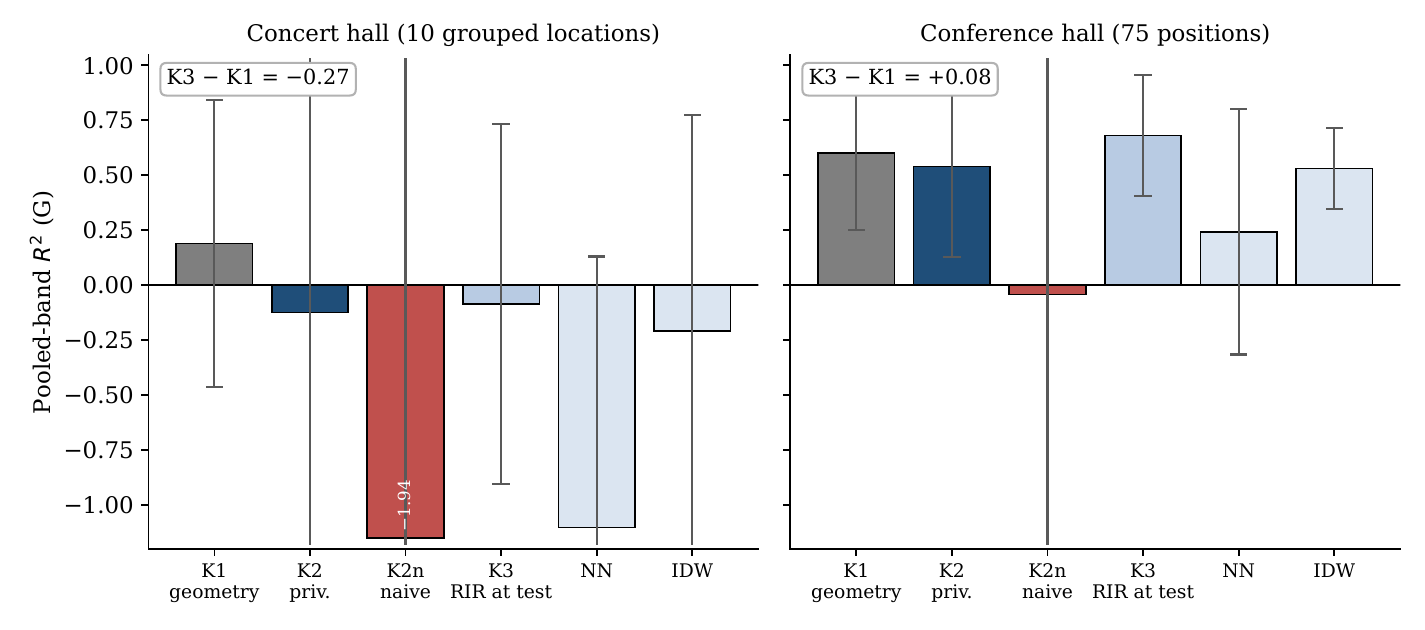}
\caption{Sound strength under the six conditions of
Table~\ref{tab:cnn}, per hall; whiskers show one fold-to-fold standard deviation. Inset: K3 relative to geometry-only K1, a small gain within fold noise with 75 positions and none with ten receiver groups.}
\label{fig:fingerprint}
\end{figure}

Two findings carry the section. First, the privileged-information gain is null on both halls: K2 does not exceed K1 (conference hall: 0.54 versus 0.60, fold standard deviation 0.4; concert hall: $-0.12$ versus 0.19, fold standard deviation 0.7--1.2), so training-time access to the impulse response did not improve the geometry-only predictor here. The privileged mode remains the deployment-consistent way to state signal-assisted unmeasured-position claims, but on these data it delivers no measurable
benefit, and that negative result is reported explicitly. Second, test-time access to the target position's own impulse response does not recover the row-based accuracy. In the conference hall (75 positions) K3 exceeds the geometry-only K1 by 0.08 (0.68 against 0.60), a difference within the fold-to-fold standard deviation, so the acoustic branch transfers at most weakly to unseen positions' responses. In the concert hall (ten receiver groups) K3 lies 0.27 below K1 and reaches $-0.09$, although the same model class reported 0.93 under row-based folds; at this position count every grouped condition lies within fold noise of zero skill for G. The reading is that the signal branch has learnt configuration-specific features of the training positions' responses, correlated with receiver identity within the campaign but not transferable, so that the row-based figure reflects condition interpolation rather than spatial prediction. Whether test-time signal access harms or merely fails to help cannot be resolved with ten receiver groups (Figure~\ref{fig:fingerprint}).

The pattern generalises to reverberation time, included because T30 is
both the field's historically central target and the parameter for
which the fused network places the greatest weight on the acoustic
branch, making it the strongest candidate for a privileged-information
gain. The gain is again null (K2 relative to K1: $-0.04$ in the concert
hall, $0.00$ in the conference hall), and test-time signal access again yields at most a marginal change (K3 relative to K1: $+0.08$ with ten receiver groups, $+0.12$ with 75 positions, each within one fold standard deviation), so the conclusion is not specific to one parameter. The
strong T30 interpolants in the concert hall (pooled-band 0.86, above
every network condition) reflect the band-profile effect noted in the
caption of Table~\ref{tab:cnn} together with the spatial smoothness of
reverberation time in a small shoebox volume.

For clarity (C80), the baselines dominate every network
condition on both halls (conference hall: inverse-distance weighting
0.43 against a best network condition of 0.33), anticipating the
convergence result below.

\subsection{Cross-model convergence under the deployment-consistent protocol}
\label{sec:convergence}

Under grouped splits with class-I4 inputs, the three model families
arrive at comparable, modest accuracy for sound strength on the
conference hall: Random Forest 0.60 (band mean), the network's spatial branch 0.60 (pooled bands), and inverse-distance weighting 0.53, against previously reported row-based figures of 0.86 to 0.93 for the learnt models. For the clarity and temporal parameters the ordering partially
inverts: inverse-distance weighting matches or exceeds both learnt
models (Section~\ref{sec:density}). The spread between model families under this protocol (below 0.1 in $R^2$) is several times smaller than the spread between protocols for a fixed
model (0.25 for G and up to 0.75 for C80 in Table~\ref{tab:ablation}). On these data,
the protocol, not the model class, is the first-order determinant of
the reported number.

\subsection{Held-out-position density curves}
\label{sec:density}

Finally, the spatial-sampling question of the companion paper, namely how many measurement positions suffice and at what density, whose metrological ancestry~\cite{witew2010,witew2017} long predates learnt predictors, is re-posed under the deployment-consistent protocol: models are trained on $n$ receiver
positions and evaluated on all rows of the remaining $75-n$ positions,
with GEO-ENV features, twenty random position draws per density from the 75 omnidirectional positions (for LF and IACC, from their own 35 and 28 receiver positions, which admit only the fifteen-position draw), and
inverse-distance weighting computed on identical splits. On the full grid neighbouring receivers are 1.1\,m apart in the median; the median distance from a held-out receiver to its nearest training position is 1.8\,m at $n=15$ (largest 4.2\,m), 1.3\,m at $n=30$ and 1.2\,m at $n=60$, so increasing $n$ also shortens the interpolation distances.
Table~\ref{tab:density} summarises the core parameters.

\begin{table}[t]
\centering
\caption{Held-out-position density experiment, conference hall: mean $R^2$ over twenty trials and octave bands for Random Forest (RF) and inverse-distance weighting (IDW) trained on $n$ positions and evaluated on the remaining $75-n$; the 75-position column is the grouped five-fold reference of Table~\ref{tab:ablation}.}
\label{tab:density}
\begin{tabular}{lcccccccc}
\hline
 & \multicolumn{2}{c}{$n=15$} & \multicolumn{2}{c}{$n=30$} &
   \multicolumn{2}{c}{$n=60$} & RF@75 \\
Parameter & RF & IDW & RF & IDW & RF & IDW & \\
\hline
G   & 0.56 & 0.44 & 0.58 & 0.51 & 0.60 & 0.54 & 0.60 \\
T30 & $-0.07$ & $-0.03$ & 0.06 & 0.00 & 0.26 & $-0.09$ & 0.26 \\
Ts  & 0.18 & 0.19 & 0.24 & 0.28 & 0.31 & 0.32 & 0.22 \\
C80 & 0.10 & 0.12 & 0.15 & 0.20 & 0.17 & 0.23 & 0.13 \\
D50 & 0.15 & 0.17 & 0.20 & 0.26 & 0.23 & 0.29 & 0.16 \\
EDT & 0.07 & 0.06 & 0.11 & 0.14 & 0.14 & 0.19 & 0.09 \\
\hline
\end{tabular}
\end{table}

\begin{figure}[t]
\centering
\includegraphics[width=0.95\linewidth]{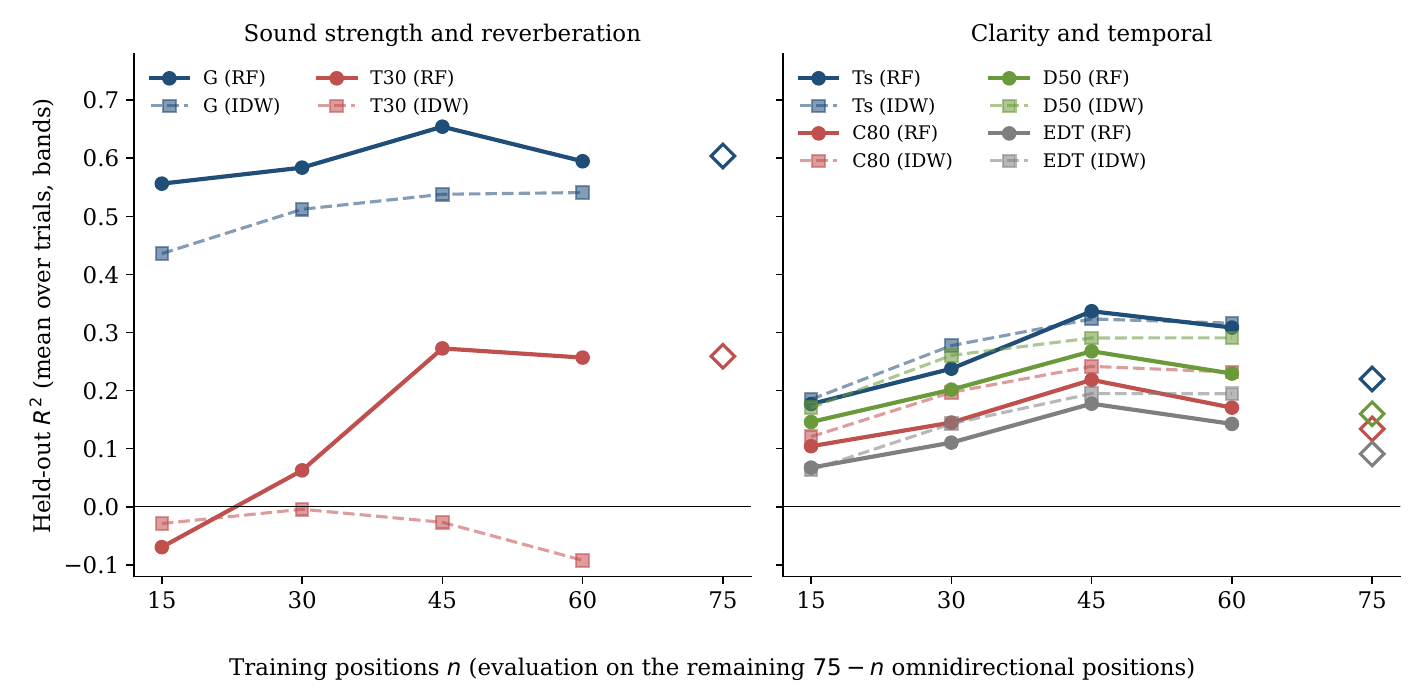}
\caption{Held-out-position density curves for the conditions of Table~\ref{tab:density}: Random Forest (solid) against inverse-distance weighting (dashed); fifteen evaluation positions remain at $n=60$. Open diamonds at $n=75$ mark the grouped five-fold reference of Table~\ref{tab:ablation}.}
\label{fig:density}
\end{figure}

Three deployment-consistent density statements replace the earlier ones. Sound
strength is the only parameter that is both moderately predictable and
robust to sparsity: it reaches 0.56 already at fifteen
positions (5.7\% of seats) and gains little thereafter, and the learnt
model outperforms interpolation for it at every density. Reverberation time is
the only parameter whose accuracy scales visibly with density ($-0.07$ at fifteen positions to 0.26 at sixty positions, matching the full-grid reference), and the learnt model
beats interpolation from thirty positions upward. For the clarity and temporal parameters, neither statement holds: accuracy remains modest at every density (the best clarity-class value is 0.34, for Ts at forty-five positions, Figure~\ref{fig:density}), and inverse-distance weighting matches (to within 0.02) or exceeds the Random Forest at every density. Two conclusions of different kinds follow and should not be conflated. As practical guidance for present-day campaigns, no density of the kind studied here explains more than about a third of the seat-to-seat variation of these parameters; whether the residual error is acceptable depends on an absolute tolerance that this paper does not set. As a statement about the prediction problem, however, this is a limit of the representations tested, not a verdict on the task: Section~\ref{sec:discussion} argues that the failure mode is representational, and that closing it is an open research question rather than a closed road. Of the
remaining parameters, the bass ratio and the initial time delay gap
reach at most 0.20 and 0.11 respectively for either model; at the fifteen-position draw that their receiver sets admit, the learnt model leads for the interaural cross-correlation (0.35 against 0.18) and neither model exceeds 0.13 for the lateral fraction. The
sixty-position values agree to within 0.1 with the 75-position grouped
reference, which corroborates the two independently implemented
harnesses (Figure~\ref{fig:density}).

 This pattern is consistent with the distinct spatial behaviour of the quantities in question: sound strength and reverberation time follow general trends, with source distance and overall absorption respectively, that a few positions constrain, whereas the clarity and temporal parameters vary from seat to seat on scales that the geometry and environmental descriptors used here cannot resolve (Section~\ref{sec:discussion}).

\section{Discussion}
\label{sec:discussion}

\subsection{The physical reading of the reordered difficulty}

Under the deployment-consistent protocol the apparent difficulty of parameter classes inverts relative to the earlier reading, and the inversion is
consistent with expectations regarding room acoustics, although the analysis does not single out one mechanism. The spatial variation of a room-acoustic quantity encompasses general trends in the source--receiver relation (distance, the balance between direct and reflected sound), systematic changes in individual reflection paths as the receiver moves, and smaller-scale interference and modal fluctuations. The relative influence of these factors varies with the quantity, the frequency band and the room; the measurement-uncertainty literature~\cite{pelorson1992,devries2001,witew2013} documents the resulting seat-to-seat variation without attributing it to a single cause. Sound strength exhibits a distinct source-distance trend, while reverberation time is largely determined by overall absorption and volume; both therefore possess components that a limited number of positions constrain, and they are the only parameters for which the learnt models offer a measurable advantage over interpolation. As far as reverberation time is concerned, some of the variation between individual positions reflects deviations from the ideal diffuse field and the uncertainty in the evaluation of the decay rather than a field at seat level; consequently, a trained T30 model provides more insight into the variation across the hall than into individual seats. The spatial variation of the clarity and temporal parameters (C80, D50, Ts, EDT) is not captured by the geometric descriptors used here (coordinates, distances and angles, without taking into account the geometry of walls, ceilings or reflectors); the analysis cannot attribute the unexplained variation to any one of the three contributors. The results indicate that this representation is insufficient, not that individual early reflections are the cause. The practical consequence for
measurement-campaign design is that guidance should be issued per
parameter class: sparse geometric prediction is supported here for sound
strength, density buys accuracy for reverberation time, and for the clarity class the representations tested here leave most of the seat-to-seat variation unexplained. This restates, on deployment-consistent terms, the per-class message of the companion density study~\cite{companion}. Whether the observed ceiling is
representational or informational is an open question; the failure mode
identified here, seat-to-seat variation unresolved by seat-level geometric descriptors, points to the former, and richer geometric encodings or physics-informed representations are the candidate approaches for testing it; physics-informed reconstruction of impulse responses from sparse measurements~\cite{borreljensen2021,karakonstantis2024} and operator learning for acoustic wave fields~\cite{middleton2025} are already moving in this direction.

\subsection{What condition interpolation remains good for}

Nothing in this paper implies that the row-based figures are
incorrect; they are correct answers to a different question, and that
question has operational value. A venue operator who has measured a set
of positions and needs parameter estimates at those same positions
under an unvisited temperature, occupancy level, or seating arrangement
is asking precisely the condition-interpolation question, and the reported accuracy (band means of 0.67 to 0.88 for the core parameters)
is the relevant figure for that use. The error the taxonomy targets is
not the number but the label attached to it: reporting a
condition-interpolation score in support of an unmeasured-position
claim. 

\subsection{Implications for system claims and practice}

Systems whose claims include inference without a sound signal, as in
the author's patent applications, make an unmeasured-position claim that is
deployment-consistent by construction (Section~\ref{sec:taxonomy-inputs}); the switchable-inference architecture of
Section~\ref{sec:models} is what allows such claims to be tested rather
than merely asserted. Two reporting duties follow. The first is that
the privileged-information gain, the difference between the
geometry-only predictor trained with and without training-time signal
access, should be reported even when it is null, as it was here on both
halls; a capability claim and a benefit claim are distinct. The second
is that test-time signal access should not be presented as evidence of
spatial generalisation without a grouped evaluation, since
Section~\ref{sec:cnn} shows that the same input reproduces the row-based accuracy without transferring to unseen positions. For consultancy practice, the deployment-consistent numbers currently support geometry-only screening
of sound strength (and, with denser grids, reverberation time), while
blind in-situ estimation~\cite{eaton2016,zhang2024} remains the
appropriate tool where a signal at the position of interest is
available; the two approaches answer complementary operational
questions.

\subsection{Relation to accuracies reported in the literature}
\label{sec:litcompare}

The protocol taxonomy offers a reading grid for the accuracy figures reported in the data-driven room-acoustics literature, although published protocol descriptions do not always permit a definitive classification. Room-level prediction, in the tradition of the early neural-network studies~\cite{nannariello1999,nannariello2001}, generalises across rooms, and blind estimation from reverberant signals~\cite{eaton2016,deng2020,duangpummet2022,zhang2024} legitimately uses a recording made at the position of interest; neither is affected by the position-memorisation channel analysed here. Concerns regarding position-level spatial prediction within a venue~\cite{falconperez2018,yu2021,bakos2025} arise wherever the validation splits are row-based, derived from repeated or augmented measurements, or the inputs include position-identifying channels. In such cases the mechanism described in Section~\ref{sec:theory} predicts inflated figures, the within-venue counterpart of the performance drop reported when large-scale ecological maps were re-validated spatially~\cite{ploton2020}. Consequently, published accuracy figures for this family of methods must be interpreted in conjunction with their split design and the inputs declared for inference. 

\subsection{Limitations}
\label{sec:limitations}

The evidence base is one multi-condition campaign in two halls; the
effect sizes are venue-specific and await cross-venue replication, for
which the audited third-hall campaign of~\cite{nvw2026}, with its prospectively pre-assigned held-out positions, is a natural candidate. Measurement-side uncertainty sources, such as source orientation and directivity~\cite{sanmartin2007}, act in addition to the protocol effects studied here and are treated in the metrological literature.
The network conditions were run for sound strength, clarity, reverberation time, and early decay time; the remaining parameters were not required for the protocol argument. Two
aggregation conventions (band pooling for the network, band averaging
for the forest) coexist in Section~\ref{sec:experiments} and make
cross-family comparisons indicative rather than exact. In the
hall with ten receiver groups, grouped five-fold estimates carry large fold
variance (1.2 in $R^2$ for sound strength under the privileged condition), which is itself part of the finding: below some position
count, spatial-prediction claims cannot be evaluated meaningfully at
all. Finally, this paper deliberately does not re-derive
perceptual (JND-referenced) error statements under the deployment-consistent protocol;
that analysis belongs with the revised density guidance
of~\cite{companion}.

\section{A reporting checklist}
\label{sec:checklist}

The findings condense into six reporting practices for data-driven room acoustics, in the same spirit in which the metrological literature has long attached explicit uncertainty statements to measured and simulated room-acoustic quantities~\cite{pelorson1992,vorlander2013}. The checklist was applied prospectively in the third-hall
campaign of~\cite{nvw2026} at no additional measurement cost.

\begin{enumerate}
\item \textbf{State the physical prediction target (held-out receiver, source, or pair) and declare the unit of generalisation}, group the folds
accordingly, and state which claims each split supports.
\item \textbf{List every inference-time input} and certify that each
exists at a genuinely unmeasured position; the target's impulse
response, co-located parameters, and position identifiers do not.
\item \textbf{Report a simple spatial baseline} (nearest neighbour or
inverse-distance weighting) on identical splits.
\item \textbf{Report per-parameter and per-band results}; averaged
scores conceal class-dependent behaviour that reverses conclusions.
\item \textbf{Separate condition interpolation from spatial
prediction}, each with its own protocol and number.
\item \textbf{Release split definitions, seeds, and evaluation code},
so that the protocol itself is reproducible.
\end{enumerate}

\section{Conclusions}
\label{sec:conclusions}

This paper asked what data-driven room-acoustic models actually see,
during validation and at inference, and answered by controlled ablation
on a single two-hall campaign. Four results carry the argument. First,
a factorial ablation of the companion Random Forest pipeline showed that
its reported accuracy is governed by two protocol choices: with
row-based folds and measured-at-test inputs the reported band means
(0.67 to 0.88) are reproduced by construction, while grouped folds with
geometry-and-environment inputs leave sound strength at 0.60,
reverberation time at 0.26, and the clarity class at 0.09 to 0.22.
Second, test-time access to the target's own impulse response, which reproduces the high row-based accuracy, was shown not to transfer to unseen positions in a small-position hall, so that the row-based figure reflects condition interpolation rather than spatial prediction, while the privileged-information gain of
training-time signal access was null on both halls. Third, under the deployment-consistent protocol the spread between the three model families, below 0.1, is several times smaller than the spread between protocols for a fixed model, while simple interpolation matches or exceeds the learnt models for the clarity class; the protocol, not the model, is the first-order determinant of the reported number. Fourth, a held-out-position density
experiment restated measurement guidance on deployment-consistent terms: sound strength is
predictable and sparsity-robust, reverberation time scales with density, and the clarity class remains largely unexplained under the representations tested. Condition interpolation remains valuable at measured positions; the requirement, operationalised by the checklist of Section~\ref{sec:checklist}, is that the two claims be stated separately, each under its own protocol.

\section*{Acknowledgments}

{\sloppy This work was supported by the Scientific and Technological Research Council of T\"{u}rkiye (T\"{U}B\.{I}TAK) under Grant 123M884.\par}

\section*{Author declarations}
\textbf{Conflict of interest.} 
The author is the inventor of two patent applications related to the systems analysed here (TR 2026/001430 and TR 2026/001431, under examination at the Turkish Patent and Trademark Office) and declares no other conflicts of interest.

\section*{Data availability}

Evaluation scripts, split definitions, seeds and per-fold results for all tables and figures are openly available in Zenodo at \url{https://doi.org/10.5281/zenodo.21453784}; the measurement data are available from the author upon reasonable request.



\begin{thebibliography}{34}

\bibitem{iso2009} ISO 3382-1:2009: Acoustics -- Measurement of room acoustic parameters -- Part 1: Performance spaces. ISO, Geneva, 2009.

\bibitem{nannariello1999} J. Nannariello, F. Fricke: The prediction of reverberation time using neural network analysis. Appl. Acoust. 58, 3 (1999) 305--325.

\bibitem{nannariello2001} J. Nannariello, F. Fricke: The use of neural network analysis to predict the acoustic performance of large rooms. {P}art {I}: Predictions of the parameter {G} utilising numerical simulations. Appl. Acoust. 62, 8 (2001) 917--950.

\bibitem{nannariello2001b} J. Nannariello, F. Fricke: The use of neural network analysis to predict the acoustic performance of large rooms. {P}art {II}: Predictions of the acoustical attributes of concert halls utilising measured data. Appl. Acoust. 62, 8 (2001) 951--977.

\bibitem{falconperez2018} R. Falc\'{o}n P\'{e}rez: Machine-learning-based estimation of room acoustic parameters. MSc thesis, Aalto University, 2018.

\bibitem{bianco2019} M.J. Bianco, P. Gerstoft, J. Traer, E. Ozanich, M.A. Roch, S. Gannot, C. Deledalle: Machine learning in acoustics: Theory and applications. J. Acoust. Soc. Am. 146, 5 (2019) 3590--3628.

\bibitem{eaton2016} J. Eaton, N.D. Gaubitch, A.H. Moore, P.A. Naylor: Estimation of room acoustic parameters: The {ACE} {C}hallenge. IEEE/ACM Trans. Audio Speech Lang. Process. 24, 10 (2016) 1681--1693.

\bibitem{deng2020} S. Deng, W. Mack, E.A.P. Habets: Online blind reverberation time estimation using {CRNN}s, in Proc. Interspeech, 2020.

\bibitem{duangpummet2022} S. Duangpummet, J. Karnjana, W. Kongprawechnon, M. Unoki: Blind estimation of speech transmission index and room acoustic parameters based on the extended model of room impulse response. Appl. Acoust. 185 (2022) 108372.

\bibitem{zhang2024} Y. Zhang, J. Sang, C. Zheng, X. Li: A denoising-aided multi-task learning method for blind estimation of reverberation time. Measurement 231 (2024) 114568.

\bibitem{yu2021} W. Yu, W.B. Kleijn: Room acoustical parameter estimation from room impulse responses using deep neural networks. IEEE/ACM Trans. Audio Speech Lang. Process. 29 (2021) 436--447.

\bibitem{yeh2021} C. Yeh, Y. Tsay: Using machine learning to predict indoor acoustic indicators of multi-functional activity centers. Applied Sciences 11, 12 (2021) 5641.

\bibitem{bakos2025} B. Bakos, G. Hidy, B. Csan{\'a}dy, C. Huszty, A. Luk{\'a}cs: Estimating room acoustic descriptors from bag-of-vectors representation with transformers. Eng. Appl. Artif. Intell. 145 (2025) 110183.

\bibitem{vanwaterschoot2025} T. van Waterschoot: Deep, data-driven modeling of room acoustics: literature review and research perspectives. arXiv:2504.16289 (2025).

\bibitem{companion} A. Oktav: Spatial sampling requirements for data-driven prediction of room acoustic parameters: a within-venue sub-sampling experiment. Build. Acoust. (2026), \url{https://doi.org/10.1177/1351010X261487597}.

\bibitem{kaufman2012} S. Kaufman, S. Rosset, C. Perlich, O. Stitelman: Leakage in data mining: Formulation, detection, and avoidance. ACM Trans. Knowl. Discov. Data 6, 4 (2012) 1--21.

\bibitem{kapoor2023} S. Kapoor, A. Narayanan: Leakage and the reproducibility crisis in machine-learning-based science. Patterns 4, 9 (2023) 100804.

\bibitem{saeb2017}
S. Saeb, L. Lonini, A. Jayaraman, D.C. Mohr, K.P. Kording: The need to
approximate the use-case in clinical machine learning. GigaScience 6,
5 (2017) gix019.

\bibitem{roberts2017} D.R. Roberts, V. Bahn, S. Ciuti, M.S. Boyce, J. Elith, G. Guillera-Arroita, S. Hauenstein, J.J. Lahoz-Monfort, B. Schr\"{o}der, W. Thuiller, D.I. Warton, B.A. Wintle, F. Hartig, C.F. Dormann: Cross-validation strategies for data with temporal, spatial, hierarchical, or phylogenetic structure. Ecography 40, 8 (2017) 913--929.

\bibitem{ploton2020}
P. Ploton, F. Mortier, M. R\'ejou-M\'echain, N. Barbier, N. Picard,
V. Rossi, C. Dormann, G. Cornu, G. Viennois, N. Bayol, A. Lyapustin,
S. Gourlet-Fleury, R. P\'elissier: Spatial validation reveals poor
predictive performance of large-scale ecological mapping models.
Nat. Commun. 11 (2020) 4540.

\bibitem{nvw2026} A. Oktav: Measurement data quality in room acoustics: a systematic audit framework for field campaign errors and their quantified impact on machine learning prediction performance. Noise Vib. Worldw. (2026), \url{https://doi.org/10.1177/09574565261477338}.

\bibitem{farina2000} A. Farina: Simultaneous measurement of impulse response and distortion with a swept-sine technique, in 108th AES Convention, Paris, 2000.

\bibitem{schroeder1965} M.R. Schroeder: New method of measuring reverberation time. J. Acoust. Soc. Am. 37, 3 (1965) 409--412.

\bibitem{vapnik2009} V. Vapnik, A. Vashist: A new learning paradigm: Learning using privileged information. Neural Netw. 22, 5--6 (2009) 544--557.

\bibitem{breiman2001} L. Breiman: Random forests. Mach. Learn. 45, 1 (2001) 5--32.

\bibitem{lecun2015} Y. LeCun, Y. Bengio, G. Hinton: Deep learning. Nature 521, 7553 (2015) 436--444.

\bibitem{witew2010} I.B. Witew, P. Dietrich, D. de Vries, M. Vorl\"{a}nder: Uncertainty of room acoustical measurements: How many measurement positions are necessary to describe the conditions in auditoria?, in Proc. ISRA, Melbourne, 2010.

\bibitem{witew2017} I.B. Witew, M. Vorl\"{a}nder, N. Xiang: Sampling the sound field in auditoria using large natural-scale array measurements. J. Acoust. Soc. Am. 141, 3 (2017) EL300.

\bibitem{pelorson1992} X. Pelorson, J. Vian, J. Polack: On the variability of room acoustical parameters: Reproducibility and statistical validity. Appl. Acoust. 37, 3 (1992) 175--198.

\bibitem{devries2001} D. de Vries, E.M. Hulsebos, J. Baan: Spatial fluctuations in measures for spaciousness. J. Acoust. Soc. Am. 110, 2 (2001) 947--954.

\bibitem{witew2013} I.B. Witew, P. Dietrich, S. Pelzer, M. Vorl\"{a}nder: Comparison of strategies to model spatial fluctuations of room acoustic single number quantities. Build. Acoust. 20, 4 (2013) 323.

\bibitem{borreljensen2021} N. Borrel-Jensen, A.P. Engsig-Karup, C. Jeong: Physics-informed neural networks for one-dimensional sound field predictions with parameterized sources and impedance boundaries. JASA Express Letters 1, 12 (2021) 122402.

\bibitem{karakonstantis2024} X. Karakonstantis, D. Caviedes-Nozal, A. Richard, E. Fernandez-Grande: Room impulse response reconstruction with physics-informed deep learning. J. Acoust. Soc. Am. 155, 2 (2024) 1048--1059.

\bibitem{middleton2025} M. Middleton, D.T. Murphy, L. Savioja: Modelling of superposition in {2D} linear acoustic wave problems using {Fourier} neural operator networks. Acta Acustica 9 (2025) 20.

\bibitem{sanmartin2007} R. San Mart\'{i}n, I.B. Witew, M. Arana, M. Vorl\"{a}nder: Influence of the source orientation on the measurement of acoustic parameters. Acta Acust. united Ac. 93, 3 (2007) 387--397.

\bibitem{vorlander2013} M. Vorl\"{a}nder: Computer simulations in room acoustics: Concepts and uncertainties. J. Acoust. Soc. Am. 133, 3 (2013) 1203--1213.

\end{thebibliography}
\end{document}